\documentclass[12pt]{iopart} 

\usepackage{subfig}

\expandafter\let\csname equation*\endcsname\relax
\expandafter\let\csname endequation*\endcsname\relax
\usepackage{amsmath}
\usepackage{graphicx}
\newtheorem{proposition}{Proposition}

\usepackage{iopams,cite}

\begin{document}

\title[]{Deconfined quantum criticality and generalised
exclusion statistics in a non-hermitian BCS model
}

\author{Jon Links, Amir Moghaddam, and Yao-Zhong Zhang}

\address{Centre for Mathematical Physics, School of Mathematics and Physics, The University of Queensland 4072,
 Australia}

\eads{\mailto{jrl@maths.uq.edu.au}}

\begin{abstract}
We present a pairing Hamiltonian of the Bardeen-Cooper-Schrieffer form which exhibits two quantum critical lines of deconfined excitations. This conclusion is drawn using the exact Bethe ansatz equations of the model which admit a class of analytic solutions. The deconfined excitations obey generalised exclusion statistics.  A notable property of the Hamiltonian is that it is non-hermitian. Although it does not have a real spectrum for all choices of coupling parameters, we provide a rigorous argument to establish that real spectra occur on the critical lines. The critical lines are found to be invariant under a renormalisation group map.
\end{abstract}


%\maketitle

\section{Introduction} Characterising quantum states of matter has been a subject of fervent activity for some years now, using a variety of techniques. Terms such as order parameter, renormalisation group, excitation gap, entanglement, fidelity, topological invariant, and conformal invariance are commonly found in the identification of quantum critical (or quantum phase transition) points between distinct quantum phases. In \cite{svbsf04,sbsvf04} it was proposed that quantum criticality may be identified by deconfined (or emergent) excitations at the critical point, which are not found in phases adjacent to the critical point. Here we present a pairing Hamiltonian of the Bardeen-Cooper-Schrieffer (BCS) form \cite{bcs57} which provides an example of this phenomenon through two lines of critical points.       
This model has five properties which, in combination, give it a unique profile: 1) the model admits an exact Bethe ansatz solution 
whereby the energy spectrum associated to the deconfined excitations can be calculated analytically; 2) the deconfined excitations exist for finite-sized systems, not just in the thermodynamic limit; 3) the deconfined excitations obey generalised exclusion statistics;  4) the Hamiltonian is non-hermitian but has a real spectrum on the critical lines; 5) the spectrum of deconfined excitations is invariant under a renormalisation group map.    

\section{The Hamiltonian}  
The general form for a reduced BCS Hamiltonian as originally discussed in \cite{bcs57} is given by 
\begin{align} H_{\rm{BCS}}=\sum_{j=1}^{L}\epsilon_j n_j
-\sum_{j,k=1}^{L}G_{jk} 
c_{k+}^{\dagger}c_{k-}^{\dagger}c_{j-} c_{j+}. \label{h_bcs} 
\end{align}
Here, $j=1,\dots,{L}$ enumerates doubly-degenerate, single-particle
energy levels with energy $\epsilon_j$  for
level $j$. The operators $c_{j\pm} ,\,c^{\dagger}_{j\pm}$ are annihilation
and creation operators for fermions at level $j$, and $n_j=c^\dagger_{j+}c_{j+}
 + \ c^\dagger_{j-}c_{j-} $ are 
fermion number operators. The labels $\pm$
refer
to pairs of time-reversed states. Throughout we will work with a {\it picket fence} model whereby the $\epsilon_j$ are uniformly and symmetrically distributed around zero. In particular we choose 
\begin{align}
\epsilon_j = \left(j-\frac{L+1}{2}\right)\delta
\label{pf}
\end{align}
where the level spacing $\delta$ provides an energy scale for the system.

The study of exactly solvable cases of BCS Hamiltonians originates from the work of Richardson \cite{r63} dealing with uniform couplings $G_{jk}=G$ for all $j,k$. This case is also known as the $s$-wave pairing model. 
Our interest is in the choice  
\begin{align}
G_{jk}=\begin{cases}
  G_+ ,& j< k , \\
  \displaystyle \frac{G_++G_-}{2}, & j=k, \\
  G_- ,& j>k
\end{cases} 
\label{gees}
\end{align}
for two independent parameters $G_+$ and $G_-$. This contains the Richardson subcase when $G_+=G_-$. The instance where  $G_+$ and $G_-$ are a complex conjugate pair is known as the {\it Russian Doll} model, which has a self-adjoint Hamiltonian. It was introduced in \cite{lrs04} as an example of a many-body system admitting a cyclic renormalisation group map, motivated by the one-body model of Glazek and Wilson \cite{gw}. The Russian Doll model was shown to be exactly solvable in \cite{dl04}. Below we study (\ref{h_bcs},\ref{pf},\ref{gees}) for real-valued $G_+,\,G_->0$. It will be shown that the critical lines are given by 
\begin{align}
G_+ - G_-=\pm 2\delta.   
\label{crit}
\end{align}

An important feature of the Hamiltonian (\ref{h_bcs})
is the blocking
effect. For any unpaired fermion at level $j$, the action of
the pairing interaction is zero since only paired fermions are
scattered.
This means that the Hilbert space can be decoupled into
a product of paired and unpaired fermion states in which the
action of the Hamiltonian on the space for the unpaired fermions is
diagonal in the basis of number operator eigenstates.
In view of this property the pair number operator 
$\displaystyle N=\sum_{j=1}^L c^\dagger_{j+}c_{j+}c_{j-}^\dagger c_{j-}$
commutes with (\ref{h_bcs}) and thus provides a good quantum number. Throughout, $M$ will be used to denote the eigenvalues of the pair number operator, while $m$ will denote the eigenvalues of the total fermion number operator $n=\displaystyle \sum_{j=1}^L n_j$. 

\begin{figure} 
\centering
{\includegraphics[width=7.3cm]{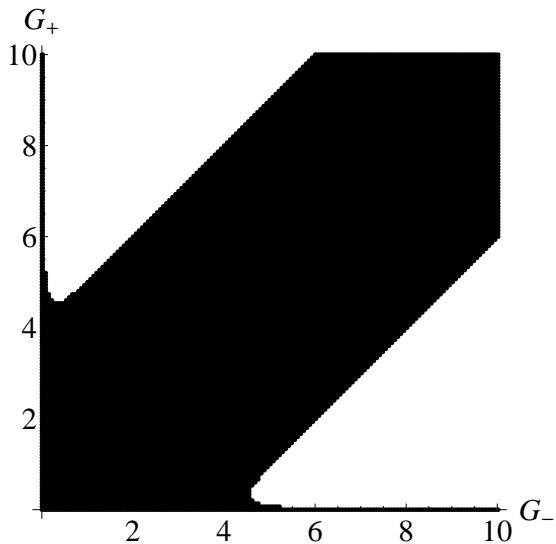}}
\caption{The shaded region depicts the values of the coupling parameters $0 \leq G_+,\,G_-\leq 10$ for which (\ref{h_bcs},\ref{pf},\ref{gees}) has real spectrum when $\delta=2$, $L=8$, $m=6$, and $M=2$, with blocked levels at $\epsilon_2=-5$ and $\epsilon_6=3$. For other parameter choices we also observe real spectra in the region between the lines (\ref{crit}).  }
\label{fig1}
\end{figure}

First we address that issue that the Hamiltonian is generally non-hermitian for real-valued $G_+,\,G_->0$, so is not guaranteed to have a real spectrum (except when $G_+=G_-$, or when either of $G_{\pm}$ is zero). The study of non-hermitian Hamiltonians with real spectra has attracted intense activity \cite{special}. Numerical diagonalisation for small system sizes shows that (\ref{h_bcs}) subject to (\ref{pf},\ref{gees}) does give rise to complex spectra for some choices of the coupling parameters, as illustrated in Fig. 1. The numerical results show that the boundaries between real and complex spectra are close to the lines given by (\ref{crit}). Next we turn to the exact solution of the Hamiltonian to show that 
(\ref{crit}) are associated with deconfined excitations obeying generalised exclusion statistics.

\section{Exact Bethe ansatz solution and generalised exclusion statistics}  
The exact solution  for (\ref{h_bcs}) subject to (\ref{gees}) was obtained using the techniques of the {\it Quantum Inverse Scattering Method} and {\it algebraic Bethe ansatz} \cite{faddeev}. In order to present the exact solution, which is adapted from \cite{dl04}, it is useful to make a change of variables. We parameterise the coupling constants $G_\pm$ through variables $\alpha$, $\eta$ such that 
\begin{align}
\alpha&=\frac{1}{2}\ln \left(\frac{G_+}{G_-}\right),\quad
\eta= \frac{1}{2}\left(G_+ - G_-\right).  
\label{change}
\end{align}   
%Inverting these relations gives
%\begin{align*}
%G_+&=\frac{2\eta e^{\alpha}}{e^\alpha - e^{-\alpha}},\\
%G_-&=\frac{2\eta e^{-\alpha}}{e^\alpha - e^{-\alpha}}
%\end{align*} 
%
For our purposes it will be sufficient to restrict the subsequent analysis to the subspace with no unpaired fermions (i.e. $M=m/2$). The exact solution for the energy spectrum on this subspace is 
\begin{align}
E=2\sum_{k=1}^M v_k
\label{nrg}
\end{align}
where the $v_k,\,k=1,...,M$, are solutions of the system of {\it Bethe ansatz equations}
\begin{align}
&e^{2\alpha} P(v_k+\eta) \prod^M_{j\neq k}(v_k-v_j-\eta)  =P(v_k) \prod^M_{j\neq k}(v_k-v_j+\eta), \quad k=1,...,M    
\label{bae}
\end{align}
with $\displaystyle P(u)=\prod_{j=1}^L(u-\epsilon_j-\eta/2)$. Each of the solutions $v_k$ may be viewed as the energy of a fermion in a Cooper pair quasiparticle. However these quasiparticles are typically bound rather than free, since a solution set for $M$ quasiparticles is not simply the union of one-body solutions due to the coupled nature of (\ref{bae}). 

In general the Eqs. (\ref{bae}) cannot be solved analytically. However  setting $\eta=\pm\delta$,  and recalling (\ref{pf}), it is seen that the polynomials $P(u)$ and $P(u+\eta)$ share a set of common roots. Specifically $P(v)=P(v+\eta)=0$ for $ v\in {\mathcal S}$ where
\begin{align}
{\mathcal S}= \{\delta(k-L/2):k=1,...,L-1 \}. 
\label{s}
\end{align}   
For this case we obtain analytic solutions of (\ref{bae}) by choosing $v_k\in {\mathcal S}$ for $k=1,...,M$, with associated energies given by (\ref{nrg}).  A remarkable property of these solutions, corresponding to the lines (\ref{crit}), is that the energies are independent of the parameter $\alpha$. This justifies identifying them with deconfined excitations which do not occur for generic values of $G_\pm$. Note however that these particular solutions which occur on the critical lines are {\it not} associated with the ground state of the model, and that there is a gap between the ground and excited states.

For these deconfined excitations a free quasiparticle interpretation {\it is} appropriate, since a solution set for $M$ particles is simply the union of one-body solutions chosen from ${\mathcal S}$. However there are still restrictions that apply to the possible choices, which leads to the picture of free quasiparticles obeying generalised exclusion statistics in the sense proposed by Haldane \cite{h91}. Partially deconfined excitations also occur, where some roots belong to ${\mathcal S}$ and others do not, but will not be discussed here. 
Next we state the main result before explaining the details behind it: 

\begin{proposition}
For the Hamiltonian (\ref{h_bcs}) subject to (\ref{pf},\ref{gees},\ref{crit}), consider a set ${\mathcal T}=\{v_1,\,v_2,\,...,v_M\}$ where each $v_j\in\,{\mathcal S}$ as given by (\ref{s}). Then ${\mathcal T}$ gives rise to an energy eigenvalue of a deconfined excitation through (\ref{nrg}) provided that for each pair $v_j,\,v_k\in\, {\mathcal T}$ we have $|v_j-v_k|> \delta$.  For given $L$ and $M\leq L/2$ the total number of deconfined excitations is  
\begin{align}
n(L,M)=\frac{(L-M)!}{(L-2M)!M!} 
\label{number}
\end{align}
and they are said to obey generalised exclusion statistics. A consequence is that deconfined excitations do not occur for filling fractions greater than 1/2.
\end{proposition} 

This result stems from certain aspects of the representation theory of the Yangian algebra $Y(gl(2))$ associated to the Lie algebra $gl(2)$ \cite{cp90}, which is the algebraic structure underpinning the exact solution of the Hamiltonian through the algebraic Bethe ansatz \cite{faddeev}. For our purposes we view this algebra as being dependent on a variable, which in the present context corresponds to $\eta$. 

The finite-dimensional, highest-weight modules of $Y(gl(2))$ are characterised by a polynomial, in an analogous way that finite-dimensional, highest-weight modules of $gl(2)$ are characterised by a highest-weight vector. Typically such a polynomial is termed a {\it Drinfeld polynomial}. Every polynomial is a Drinfeld polynomial, through its decomposition into {\it strings} \cite{cp90}, for some finite-dimensional, highest-weight module. Given a polynomial $Q(u)$ of order $L$, if $Q(u)$ and $Q(u+\eta)$ do not have common roots then the $Y(gl(2))$-module which has $Q(u)$ as its Drinfeld polynomial is irreducible of dimension $2^L$. If the roots are then varied such that a common root does occur, at say $u=w$, the module contains a non-trivial submodule. In such a case we can write $Q(u)=(u-w)(u-w-\eta)R(u)$ such that $Q(w)=Q(w+\eta)=0$. Then the module associated to $Q(u)$ contains a submodule containing a 
highest-weight state associated with the Drinfeld polynomial $R(u)$. 
   
For the case at hand with generic values of the level spacing $\delta$, the Drinfeld polynomial is the polynomial $P(u)$ which appears in (\ref{bae}) and is associated with an irreducible module. However setting $\eta=\pm\delta$, $P(u)$ and $P(u+\eta)$ have many common roots and non-trivial submodules arise. For a given index $j$, $2j\in\,{\mathbb Z}$ with $2-L\leq 2j\leq L-2$, consider the expression
$$P(u)=(u-(j-1)\delta)(u-j\delta)(u-(j+1)\delta)S(u)$$
where $S(u)$ is a polynomial of degree $L-3$.  
We set $v_1=j\delta $ and $v_2=(j-1)\delta$ such that $v_1-v_2=\delta$ and $P(v_1)=P(v_1+\delta)=P(v_2)=P(v_2+\delta)=0$ with $S(v_1)=S(v_2+\delta)\neq 0$. By the previous discussion the module associated with $P(u)$ contains a submodule with a highest-weight state associated with the Drinfeld polynomial 
$R_1(u)=(u-v_2)S(u)$. This highest-weight state has energy $E_1=2v_1$. The module associated with $P(u)$ also contains a submodule with a highest-weight state associated to the Drinfeld polynomial 
$R_2(u)=(u-(v_1+\delta))S(u)$. This highest-weight state has energy $E_2=2v_2$.    
Now $R_1(v_2)=0$ but $R_1(v_2+\delta)\neq 0$, and $R_2(v_1+\delta)= 0$ but $R_2(v_1)\neq 0$. So there is no highest-weight state to be found with energy $E_{1+2}=E_1+E_2$. For this reason we cannot take both $v_1$ and $v_2$ as elements of a solution set for (\ref{bae}). A straightforward counting argument leads to (\ref{number}). %\footnote{Using computer symbolic manipulation capabilities we have verified {\it analytically} that this generalised exclusion principle holds for all $L\leq 6$.}.  

At a mathematical level the Hamiltonian is very closely related in some respects, but quite different in others, to the Haldane-Shastry model \cite{h_s} for which $Y(gl(2))$ also plays an intimate role \cite{bghp93}. It is well-known that some distinguishing features of the Haldane-Shastry model are that the energy levels are known analytically, they contain high degeneracies, they have integer spacing (in appropriate units),  and the excitations are categorised as being {\it semionic} \cite{h91a}. This bears some similarity to the spectrum of deconfined excitations described above. The quantum Lax operator as given in \cite{bghp93} for the Haldane-Shastry model is equivalent, up to a non-unitary transformation, to the quantum Lax operator for (\ref{h_bcs}). The transfer matrix obtained by using the form of the Lax operator in \cite{bghp93} is self-adjoint on the critical lines, from which it follows that the spectrum of (\ref{h_bcs},\ref{pf},\ref{gees}) is real on the critical lines. A key difference is the inclusion of the ``twist in the boundary conditions'' parameterised by the variable $\alpha$ \cite{dl04}, which does not have an analogue in the Haldane-Shastry model.  Another contrasting feature is that the Haldane-Shastry model Hamiltonian is not derived from a transfer matrix associated to the Yangian algebra, but is more closely related to the quantum determinant \cite{wgx97}.     
    
\section{Renormalisation group map} The motivation of \cite{lrs04} to introduce the Russian Doll model was for the study of cyclic renormalisation group maps. The mathematical difference between the Russian Doll model  and the non-hermitian model considered here is simply through the change of variables  
$\eta \rightarrow i\eta, \alpha \rightarrow i\alpha$. Thus we can directly transcribe the renormalisation group map for the non-hermitian model from the results of \cite{lrs04}. For a system of $L$ levels, eliminating high magnitude energy degrees of freedom associated with $\epsilon_1$ or $\epsilon_L$ leads to a system of $L-1$ levels with renormalised coupling constants
\begin{align}
G_\pm^{(L-1)}= G_\pm^{(L)}+\frac{1}{2\delta L} G_+^{(L)}G_-^{(L)} .   
\label{rg}
\end{align} 
From (\ref{change}) and (\ref{rg}) it is seen that $\eta$ is invariant under the renormalisation group map, while $\alpha$ is not. As the spectrum of deconfined states on the critical lines is independent of $\alpha$, this spectrum is also invariant under (\ref{rg}).

\section{Mean-field analysis} 
Finally we make some comments on results obtained through a mean-field analysis, the details of which will be deferred to a later publication.
 We introduce a cut-off energy $\omega$ by setting  $\delta=2\omega/(L-1)$ such that
$\epsilon_1=-\omega$ and $\epsilon_L=\omega$. 
Letting $G_{\pm}=4\omega g_{\pm}/L$, $x=\langle{n}\rangle/(2L)$,  in the continuum limit $L\rightarrow \infty,\,G_{\pm}\rightarrow 0$ we obtain from mean-field techniques the following expressions for the chemical potential $\mu$ and gap parameter $\Delta$:
\begin{align*}
\mu=\frac{\omega(g_+^{\chi}+g_-^{\chi})(2x-1)}{g_+^{\chi}-g_-^{\chi}},  \quad
\Delta^2=\frac{16\omega^2 g_+^\chi g_-^\chi x(1-x)}{(g_+^\chi-g_-^\chi)^2},
\end{align*}
where $\chi=1/(g_+-g_-)$. Note that $\Delta^2>0$ for all $g_{\pm}$ with $0<x<1$.
The elementary excitation spectrum is found to be given by 
$\mathcal{E}(\epsilon)=\sqrt{(\epsilon-\mu)^2+\Delta^2},\, -\omega\leq \epsilon\leq \omega.$
 We obtain the gound-state energy per fermion as
\begin{align*}
e_{\rm MF}
%&=\lim_{L\rightarrow\infty} \frac{E_{\rm MF}}{2xL} \\
%&=-\frac{1}{8x\omega} \int_{-\omega}^\omega d\epsilon\,\frac{2\epsilon(\epsilon-\mu)+\Delta^2}{\sqrt{(\epsilon-\mu)^2+\Delta^2}} \\
%&= \frac{1}{2\omega} \int_{-\omega}^\omega d\epsilon\,\frac{K}{\sqrt{\epsilon^2+K}} 
%- \frac{1}{\omega}\int_{-\omega}^\omega d\epsilon\,{\sqrt{\epsilon^2+K}} \\
%&= \frac{K}{2\omega}\ln\left( \frac{\sqrt{\omega^2+K}+\omega}{\sqrt{\omega^2+K}-\omega}\right)
%-\sqrt{\omega^2+K}-\frac{K}{\omega}\sinh^{-1}\left(\frac{\omega}{\sqrt{K}}\right) \\
&=-\frac{1}{8x\omega}\left((\omega+\mu)\sqrt{(\omega-\mu)^2+\Delta^2}+(\omega-\mu)\sqrt{(\omega+\mu)^2+\Delta^2}\right). 
\end{align*}
Note that in the hermitian limit 
$g_\pm \rightarrow g$ for which $\chi\rightarrow \infty$, both  $\mu$ and $\Delta$ can be evaluated through use of  
$\displaystyle \exp(x)=\lim_{n\rightarrow \infty} \left(1+{x}/{n}\right)^n$.
In particular for half-filling $x=1/2$ we obtain $\mu=0$ and  
$\Delta={\omega}/{\sinh(1/2g)}$ 
which is in agreement with the classic $s$-wave result obtained in \cite{bcs57} (equation (2.40)).

From the above results we point out the following two observations, which pose some open questions to be addressed in future work. 
\begin{itemize}
\item[a)] In the mean-field approximation the non-hermitian Hamiltonian has real spectrum for all couplings $g_\pm$ and filling fractions $x$. It might be expected that mean-field results are exact in the thermodynamic limit (e.g. see \cite{rsd02} for when $g_+=g_-$). It would be very useful to determine whether or not the energy spectrum is real to leading order, with complex terms only appearing in lower order corrections. 
\item[b)] In terms of the mean-field variables the critical lines (\ref{crit}) correspond to $\chi=\pm 1$. Here the mean-field excitation spectrum of the Bogoliubov quasiparticles does not obviously display some subset of excitations with the property of being invariant with respect to $\alpha$. It is consequently not clear whether it is possible to reproduce generalised exclusion statistics within the  Bogoliubov quasiparticle picture. 
\end{itemize}
\ack We thank an anonymous referee for very useful comments. Jon Links  and Yao-Zhong Zhang are supported by the Australian Research Council through Discovery Projects DP110101414 and DP110103434 respectively. Amir Moghaddam is supported by an International Postgraduate Research Scholarship and a UQ Research Scholarship.


\section*{References}
\begin{thebibliography}{99}

\bibitem{svbsf04} T. Senthil, A. Vishwanath, L. Balents, S. Sachdev, and M.P.A. Fisher, Science {\bf 303}, 1490 (2004).

\bibitem{sbsvf04} T. Senthil, L. Balents, S. Sachdev, A. Vishwanath, and M.P.A. Fisher, Phys. Rev. B {\bf 70}, 144407 (2004).

\bibitem{bcs57} J. Bardeen, L.N. Cooper, and J.R. Schrieffer, Phys. Rev. {\bf 108}, 1175 (1957). 

\bibitem{r63} R.W. Richardson, Phys. Lett. {\bf 3}, 277 (1963).

\bibitem{lrs04} A. LeClair, J.M. Rom\'an, and G. Sierra, Phys. Rev. B {\bf 69}, 020505(R) (2004).

\bibitem{gw} St.D. Glazek and K.G. Wilson, Phys. Rev. Lett. {\bf 89}, 230401 (2002).

\bibitem{dl04} C. Dunning and J. Links, Nucl. Phys. B {\bf 702}, 481 (2004). 

\bibitem{special} C.M. Bender, Rep. Prog. Phys. {\bf 70}, 947 (2007); C.M. Bender, A. Fring, U. Guenther, and H.F. Jones, J. Phys. A: Math. Theor. {\bf 45}, 010201 (2012). 

\bibitem{faddeev} L.A. Takhtadzhan and L.D. Faddeev, Russ. Math. Surv. {\bf 34}, 11 (1979).

\bibitem{h91} F.D.M. Haldane, Phys. Rev. Lett. {\bf 67}, 937 (1991).

\bibitem{cp90} V. Chari and A. Pressley, L'Enseignement Math. {\bf 36}, 267 (1990).

\bibitem{h_s} F.D.M. Haldane, Phys. Rev. Lett. {\bf 60}, 635 (1988); B.S. Shastry, Phys. Rev. Lett. {\bf 60}, 639 (1988).

\bibitem{bghp93} D. Bernard, M. Gaudin, F.D.M. Haldane, and V. Pasquier, J. Phys. A: Math. Gen. {\bf 26}, 5219 (1993).

\bibitem{h91a} F.D.M. Haldane, Phys. Rev. Lett. {\bf 66}, 1529 (1991).

\bibitem{wgx97} Z.-F. Wang, M.-L. Ge, and K. Xue, J. Phys. A: Math. Gen. {\bf 30}, 5023 (1997).  

\bibitem{rsd02} J.M. Rom\'an, G. Sierra, and J. Dukelsky, Nucl. Phys. B {\bf 634}, 483 (2002).














\end{thebibliography}
\end{document}